\documentclass[twocolumn,prl,amsfonts,floatfix,superscriptaddress]{revtex4}

\ifx\pdfoutput\undefined
\usepackage[dvips]{graphicx}
\else
\usepackage[pdftex]{graphicx}

\usepackage[pdftex]{hyperref}
\fi
\usepackage{amsmath,amsfonts,amssymb,amscd,amsthm,epsf}
\usepackage{subfigure}
\usepackage{color}

\graphicspath{{figs/}}

\theoremstyle{remark}

\newcommand{\eg}{{e.g., }}
\newcommand{\Rset}{{\mathbb R}}
\newcommand{\Tset}{{\mathbb T}}
\newcommand{\nc}{\newcommand}

\nc{\figref}[1]{Fig.~\ref{fig:#1}}
\nc{\figsref}[2]{Figs.~\ref{fig:#1}-\ref{fig:#2}}
\nc{\tabref}[1]{Table~\ref{tab:#1}}
\nc{\tabsref}[2]{Tables~\ref{tab:#1}-\ref{tab:#2}}
\nc{\secref}[1]{Sec.~\ref{sec:#1}}
\nc{\secsref}[2]{Sec.~\ref{sec:#1}-Sec.~\ref{sec:#2}}
\nc{\ssecref}[1]{Sec.~\ref{ssec:#1}}
\nc{\ssecsref}[2]{Sec.~\ref{ssec:#1}-Sec.~\ref{ssec:#2}}
\nc{\eqeqref}[1]{Eq.~\eqref{eq:#1}}
\nc{\eqseqref}[2]{Eqs.~\eqref{eq:#1}-\eqref{eq:#2}}
\nc{\thmref}[1]{Theorem~\ref{theo:#1}}
\nc{\thmsref}[2]{Theorem~\ref{theo:#1}-\ref{theo:#2}}
\nc{\rcite}[1]{Ref.~\onlinecite{#1}}
\nc{\rcites}[1]{Refs.~\onlinecite{#1}}
\nc{\qtq}[1]{{\qquad\text{#1}\qquad}}
\nc{\vect}[1]{\boldsymbol{#1}}
\definecolor{changed}{rgb}{0.3,0.3,0.3}
\nc{\changed}[1]{{\bf previous version: }{\color {changed} #1}}

\begin{document}

\title{On a new fixed point of the renormalization group operator for area-preserving maps}

\author{K.~Fuchss} \affiliation{Department of Physics and Institute
for Fusion Studies, The University of Texas at Austin, Austin, TX
78712}

\author{A.~Wurm} \affiliation{Department of Physical \& Biological
Sciences, Western New England College, Springfield, MA 01119}

\author{P.J.~Morrison} \affiliation{Department of Physics and
Institute for Fusion Studies, The University of Texas at Austin,
Austin, TX 78712}

\date{\today}

\begin{abstract}
The breakup of the shearless invariant torus with winding number
$\omega=\sqrt{2}-1$ is studied numerically using Greene's residue
criterion in the standard nontwist map. The residue behavior and
parameter scaling at the breakup suggests the existence of a new
fixed point of the renormalization group operator (RGO) for
area-preserving maps. The unstable eigenvalues of the RGO at this
fixed point and the critical scaling exponents of the torus at
breakup are computed.
\end{abstract}

\maketitle


Area-preserving nontwist maps are low-dimensional models of physical
systems whose Hamiltonians locally violate a nondegeneracy condition
(see below) as described, \eg in
\rcites{del_castillo96,apte03,wurm05}. Some applications are the
study of magnetic field lines in toroidal plasma
devices\cite{stix76,oda95,balescu98,horton98,morrison00,ullmann00,petrisor03}
and stellarators\cite{davidson95,hayashi95} (plasma physics), and
traveling waves,\cite{del_castillo93} coherent structures,
 self-consistent transport\cite{del_castillo02} (fluid dynamics), and
 particle accelerators.\cite{gerasimov86}
Nontwist regions have also been shown to appear generically in the
phase space of area-preserving maps that have a tripling bifurcation
of an elliptic fixed point.\cite{dullin00,vanderweele88} Additional
references can be found in \rcites{wurm05, del_castillo96}.

Of particular interest from a physics perspective is the breakup of
invariant tori, consisting of quasiperiodic orbits with irrational
winding number,\footnote{An {\it orbit} of an area-preserving map
$M$ is a sequence of points
$\left\{\left(x_i,y_i\right)\right\}_{i=-\infty}^{\infty}$ such that
$M\left(x_i,y_i\right) = \left(x_{i+1},y_{i+1}\right)$. The {\it
winding number} $\omega$ of an orbit is defined as the limit $\omega
= \lim_{i\to\infty} (x_i/i)$, when it exists. Here the
$x$-coordinate is ``lifted'' from $\Tset$ to $\Rset$. A {\it
periodic orbit} of period $n$ is an orbit $M^n \left( x_i,
y_i\right) = \left( x_i+m, y_i\right)$, $\forall \:i$, where $m$ is
an integer. Periodic orbits have rational winding numbers
$\omega=m/n$. An {\it invariant torus} is a one-dimensional set $C$,
a curve, that is invariant under the map, $C = M(C)$. Orbits
belonging to such a torus generically have irrational winding
number.} that often correspond to transport barriers in the physical
system, i.e., their existence determines the long-time stability of
the system. In nontwist maps, the invariant tori that appear to be
the most resilient to perturbations are the so-called {\it
shearless} tori, which correspond to local extrema in the winding
number profile of the map.

Invariant tori at breakup exhibit scale invariance under specific
phase space re-scalings, which are observed to be universal for
certain classes of area-preserving maps. To interpret these results,
a renormalization group framework has been developed (see, e.g.,
\rcites{mackay83, del_castillo97,apte03,apte05a}). For twist maps,
it is well understood which fixed point, cycle, or strange attractor
of the renormalization group operator (RGO) is encountered within a
given class of maps, depending on properties of the winding number
of the critical torus (see \rcite{chandre02} for a recent review).
For nontwist maps, however, only results for the single class of
shearless critical noble tori, i.e., shearless critical tori with
winding numbers that have a continued fraction expansion tail of
1's, are known. The result reported in this letter represents the
first new fixed point for nontwist maps.

A tool for studying the breakup of a torus with given winding number
is Greene's residue criterion, originally introduced in the context
of twist maps.\cite{greene79} This method is based  on the numerical
observation that the breakup of an invariant torus with irrational
winding number $\omega$ is determined by the stability of nearby
periodic orbits. Some aspects of this criterion have been proved for
nontwist maps.\cite{delshams00}

To study the breakup, one considers a sequence of periodic orbits
with winding numbers $q_n/p_n$ converging to $\omega$,
$\lim_{n\rightarrow\infty} q_n/p_n=\omega$. The elements of the
sequence converging the fastest are the convergents of the continued
fraction expansion of $\omega$, i.e., $[n]:=q_n/p_n=[a_0, a_1,
\ldots, a_n]$, where
\begin{equation}
\label{eq:cf} \omega=[a_0, a_1, a_2, \ldots] =
  a_0 + \cfrac{1}{a_1 + \cfrac{1}{a_2 + \ldots}}.
\end{equation}
The stability of the corresponding orbits is determined by their
residues, $R_n= [2-{\rm Tr}(DM^{p_n})]/4$, where ${\rm Tr}$ is the
trace and $DM^{p_n}$ is the linearization of the $p_n$ times
iterated map $M$ about the periodic orbit: An orbit is elliptic for
$0<R_n<1$, parabolic for $R_n=0$ and $R_n=1$, and hyperbolic
otherwise. The convergence or divergence of the residue sequence
associated with the chosen periodic orbit sequence then determines
whether the torus exists or not, respectively: If the $\omega$-torus
exists, $\lim_{n\rightarrow\infty}|R_n|=0$; if the $\omega$-torus is
destroyed, $\lim_{n\rightarrow\infty}|R_n|=\infty$. At the breakup,
different scenarios can be encountered, depending on the class of
maps and winding number of the invariant torus considered.

In nontwist maps, the residue criterion was first used in
\rcite{del_castillo96} to study the breakup of the shearless torus
of inverse golden mean $1/\gamma = (\sqrt 5 -1)/2 =
[0,1,1,1,\ldots]$ winding number in the standard nontwist map.  The
residue sequence was discovered to converge to a six-cycle. Similar
studies were conducted for other noble shearless tori of winding
numbers $\omega=1/\gamma^2$ (\rcites{wurmdiss,apte03}),
$\omega=[0,2,2,1,1,1,\ldots]$ (\rcite{aptediss}), and
$\omega=[0,1,11,1,1,1,\ldots]$ (\rcite{fuchss06}), and the same
six-cycle was found for all tori in the same symmetry class.

In this letter we study the breakup of the
$\omega=\sqrt{2}-1=[0,2,2,2,2,\ldots]$ shearless torus, which is an
example of a non-noble winding number and leads to the discovery of
a  new fixed point of the renormalization group operator of
area-preserving maps. In contrast to twist maps,\cite{satija87} the
periodicity of the elements of the continued fraction expansion has
not been linked to the periodicity of the critical residue sequence
and therefore our new result would not have been predicted. The
numerical methods we use and their accuracy are discussed in
\rcites{apte03,wurm05,fuchss06,fuchssdiss}
and we refer the reader to these publications for details.\\

As our specific model we use the {\it standard nontwist map} (SNM)
$M$ as introduced in \rcite{del_castillo93},
\begin{eqnarray}
x_{i+1} & = & x_i + a \left( 1-y^2_{i+1}\right)\nonumber\\
y_{i+1} & = & y_i -b \sin\left(2\pi x_i\right)\,,
\end{eqnarray}
where $(x,y)\in \Tset\times \Rset$ are phase space coordinates and
$a,b\in\Rset$ are parameters. This map is area-preserving and
violates the {\it twist condition}, $\partial
x_{i+1}(x_i,y_i)/\partial y_i \neq 0$, along a curve in phase space.
Although the SNM is not generic due to its symmetries, it models the
essential features of nontwist systems with a local, approximately
quadratic extremum of the winding number profile.

One important characteristic of nontwist maps is the existence of
multiple orbit chains ({\it up} and {\it down} orbits) of the same
winding number, which can undergo bifurcations when the map
parameters $a$ and $b$ are changed. When two invariant tori collide,
the winding number profile shows a local extremum and the orbit at
collision is referred to as the {\em shearless} torus. For a given
winding number these collisions occur along  bifurcation curves
$b_{\omega}(a)$ in parameter space.\cite{del_castillo96}

In order to study a shearless invariant torus, its bifurcation curve
is found numerically by approximating it by the bifurcation curves,
$b_{[n]}(a)$, of nearby periodic orbits with winding numbers that
are the continued fraction convergents of $\omega$. Greene's residue
criterion can then be used to determine where on $b_{\omega}(a)$ the
shearless torus still exists: At parameter values $a$ and the best
known approximation to $b_\omega(a)$, the residues of all periodic
orbits of convergents that have not collided, here the orbits $[n]$
with even $n$, are computed. Their limiting behavior for
$n\rightarrow\infty$ reveals the status of the torus. By repeating
the procedure for various values of $a$, with alternating residue
convergence to 0 and $\infty$, the parameter values of the shearless
torus breakup, $(a_c, b_\omega(a_c) )$, can be determined
to high precision.\\

The study of the breakup of tori with non-noble winding numbers is
difficult because, due to numerical limitations, only periodic
orbits for a small number of elements of the continued fraction
expansion can be found. Therefore one cannot numerically distinguish
between a torus that has 2's as all entries in the continued
fraction expansion and one that has 2's until one reaches the
numerical limit, and then 1's for the tail (i.e., a noble number).

To make a definite prediction for the $\omega=\sqrt{2}-1$ torus we
study the breakup of a series of sixteen invariant tori $T_i$,
starting with $\omega=[0,1,1,1,\ldots]$ up to
$\omega=[0,2,\ldots,2,1,\ldots]$ with fifteen 2's in the continued
fraction expansion.

\begin{figure}[t!] \begin{center}
\includegraphics[width=3in]{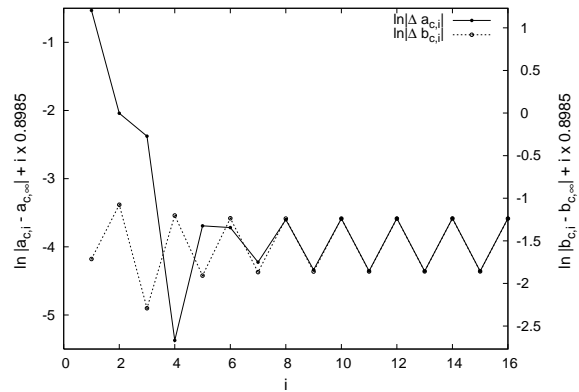}
\caption{2-cycle of $a_{c,i}$ and $b_{c,i}$ differences in
approximating the critical shearless torus $\omega=\sqrt{2}-1$ by
noble tori $T_i$. The average (negative) slope has been added to the
data.
  \label{fig:badiff}}
\end{center} \end{figure}

The breakup parameters $\left(a_{c,i},b_{c,i}\right)$ for the tori
$T_i$ are found to converge exponentially to the critical parameters
$\left(a_{c,\infty},b_{c,\infty}\right)$ of the $\omega=\sqrt{2}-1$
torus. Plotting the logarithms of the differences $\left|
a_{c,i}-a_{c,\infty}\right|$ and $\left|
b_{c,i}-b_{c,\infty}\right|$, corrected for their average slopes of
$c_a=-0.8986\pm 0.0031$ and $c_b=-0.8984\pm 0.0002$, respectively
(see \figref{badiff}), one observes a period-two oscillation. This
result should be compared to Fig. 5 in \rcite{mackay92}, where a
similar study was conducted for the (one-parameter) standard twist
map. The plot of parameter vs. $i$ showed a straight line with negative slope.\\

\begin{figure}[t!] \begin{center}
\includegraphics[width=3in]{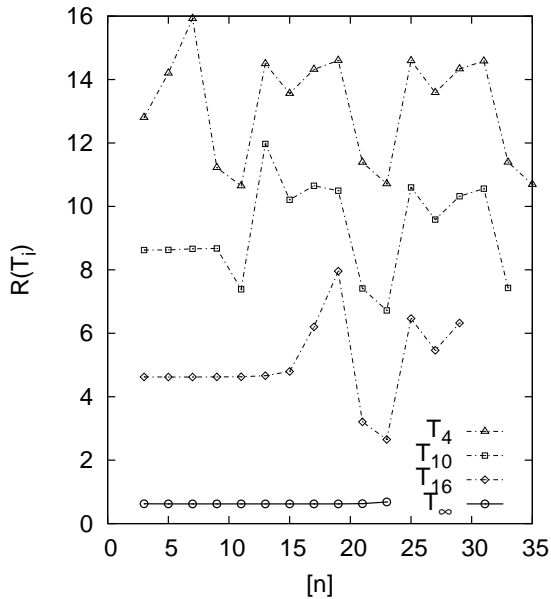}
\caption{Residue behavior at the shearless $T_i$ torus breakup for
down orbits on $s_1$. For clarity the residues of the tori have been
shifted upward by 4. (See text for definition of $T_{\infty}$.)
  \label{fig:critres}}
\end{center} \end{figure}

Figure~\ref{fig:critres} depicts the behavior of a few of the
critical residues (residues at breakup) on the $s_1$ symmetry
line\footnote{A map $M$ is called {\it reversible} if it can be
decomposed as $M=I_1\circ I_2$ with $I_i^2=0$. The fixed point sets
of $I_i$ are one-dimensional sets, called the {\it symmetry lines}
of the map. For the SNM the symmetry lines are $s_1=\{(x,y)|x=0\}$,
$s_2=\{(x,y)|x=1/2\}$, $s_3=\{(x,y)|x=a(1-y^2)/2\}$, and
$s_4=\{(x,y)|x=a(1-y^2)/2+1/2\}$.} as a function of $[n]$. The
values for different tori are shifted by 4 along the $y$-axis to
avoid overlap. The results show that the more 2's are included, the
further the familiar six-cycle for noble winding numbers gets pushed
towards higher $n$ values. We conjecture that the emerging pattern
for small $n$ values represents the critical residue pattern of the
$\omega=[0,2,2,\ldots]$ torus. The critical residues at
\[
\left(a_{c,\infty},b_{c,\infty}\right)=(0.446710414656,0.838135537624831489)
\]
along $s_1$ converge to the single value $R_c=0.621723$ for the down
orbits, and the single value $R_c=-0.909118$ for the up orbits. The
same result is found along the other symmetry lines.

As discussed in \rcites{del_castillo96,fuchss06}, close to the
critical breakup value, the $b_{[n]}(a)$ obey a scaling law
\begin{equation}
\label{eq:bscalinglaw} b_{[n]}=b_{c,\infty} + B(n)\; \delta_1 ^{-n}
\;,
\end{equation}
where $B(n)$ is numerically found to be periodic in $n$ as
$n\rightarrow \infty$ with period $4$. This period is the same as
the period of the critical fixed cycle of the renormalization group
operator ${\cal R}$ (RGO), i.e., a critical fixed point of ${\cal
R}^4$.\cite{del_castillo97}

\begin{figure}[t!] \begin{center}
\includegraphics[width=3in]{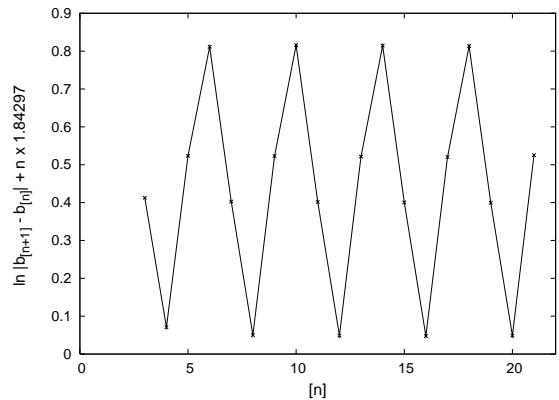}
\caption{4-cycle of $b_{[n]}$ differences in approximating the
  critical shearless $\omega=\sqrt{2}-1$ torus at
  $a_{c,\infty}=0.446710414656$ on $s_1$.
  \label{fig:bdiff}}
\end{center} \end{figure}

The scaling can be observed by plotting
\[
\label{eq:bscalinglog} \ln \left( b_{[n+1]}-b_{[n]} \right) =
\tilde{B}(n) -n\ln \delta_1 \;,
\]
where $\tilde{B}(n)=\ln(B(n+1)/\delta_1 -B(n))$ is also periodic in
$n$. This is shown (on $s_1$) in \figref{bdiff}, where for clarity
only the offsets of $\ln \left( b_{[n+1]}-b_{[n]} \right)$ about the
average slope are shown.  The slope was calculated from the last 16
difference values by averaging the last 12 slopes $\left[
\ln(b_{[n+5]}-b_{[n+4]})- \ln(b_{[n+1]}-b_{[n]}) \right] /12$, with
$n=6,\ldots,17$.
The periodicity of $\tilde{B}(n)$ makes it possible to obtain a
better approximation for $b_{c,\infty}$ (see \rcite{del_castillo96})
\[
b_{c,\infty}\approx
b_{[22]}+\frac{\left(b_{[22]}-b_{[18]}\right)\;\left(b_{[22]}-b_{[21]}\right)}{\left(b_{[19]}-b_{[18]}\right)
-\left(b_{[22]}-b_{[21]}\right)}\, .
\]
This value is used to find the critical residue pattern labeled
$T_{\infty}$ in \figref{critres}. Compared to the critical residue
six-cycle for noble winding numbers,\cite{del_castillo96} the
four-cycle of $b_{[n]}$ differences indicates that the
$\omega=\sqrt{2}-1$ breakup exhibits a residue two-cycle along each
symmetry line (only half of the periodic orbits exist at breakup).
As discussed above, in each two cycle, the numerical values of the
residues are found to be the same.

\begin{figure}[t!] \begin{center}
\includegraphics[width=3in]{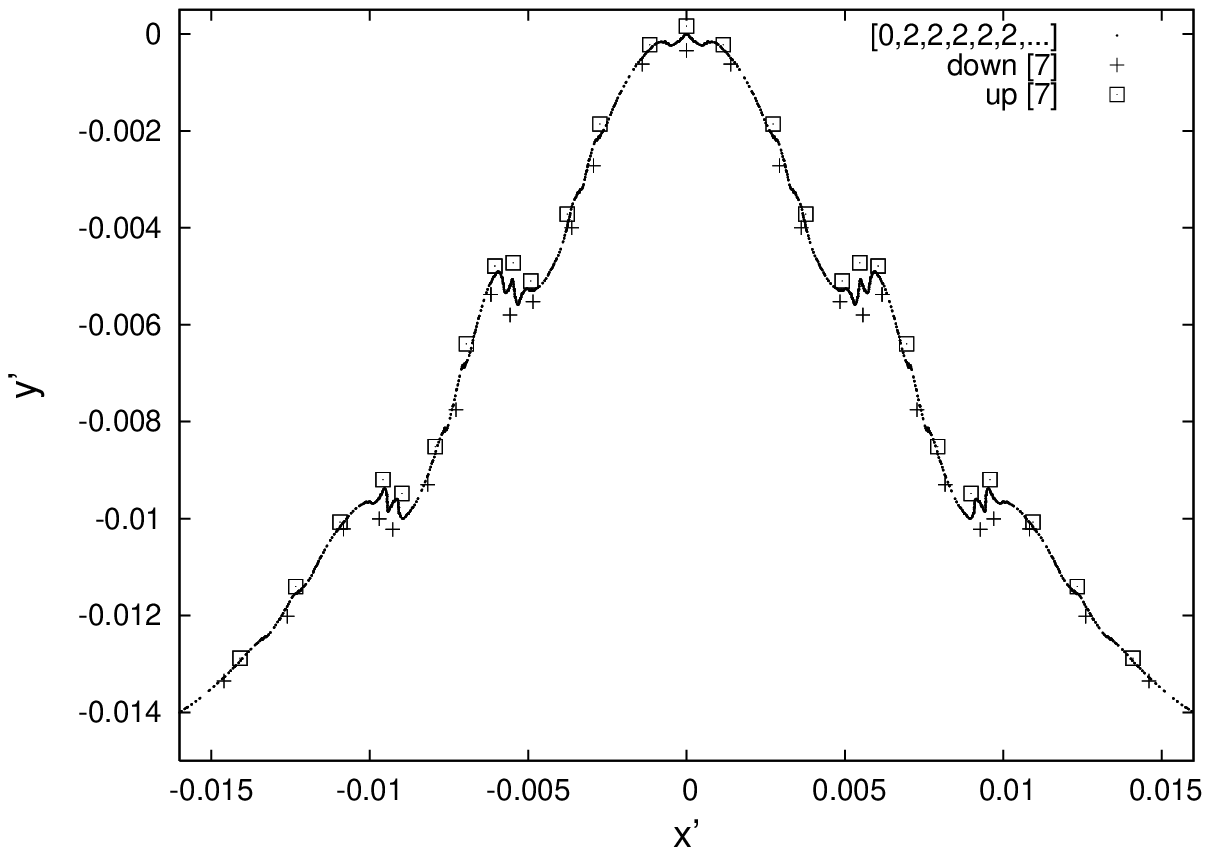}\\
\includegraphics[width=3in]{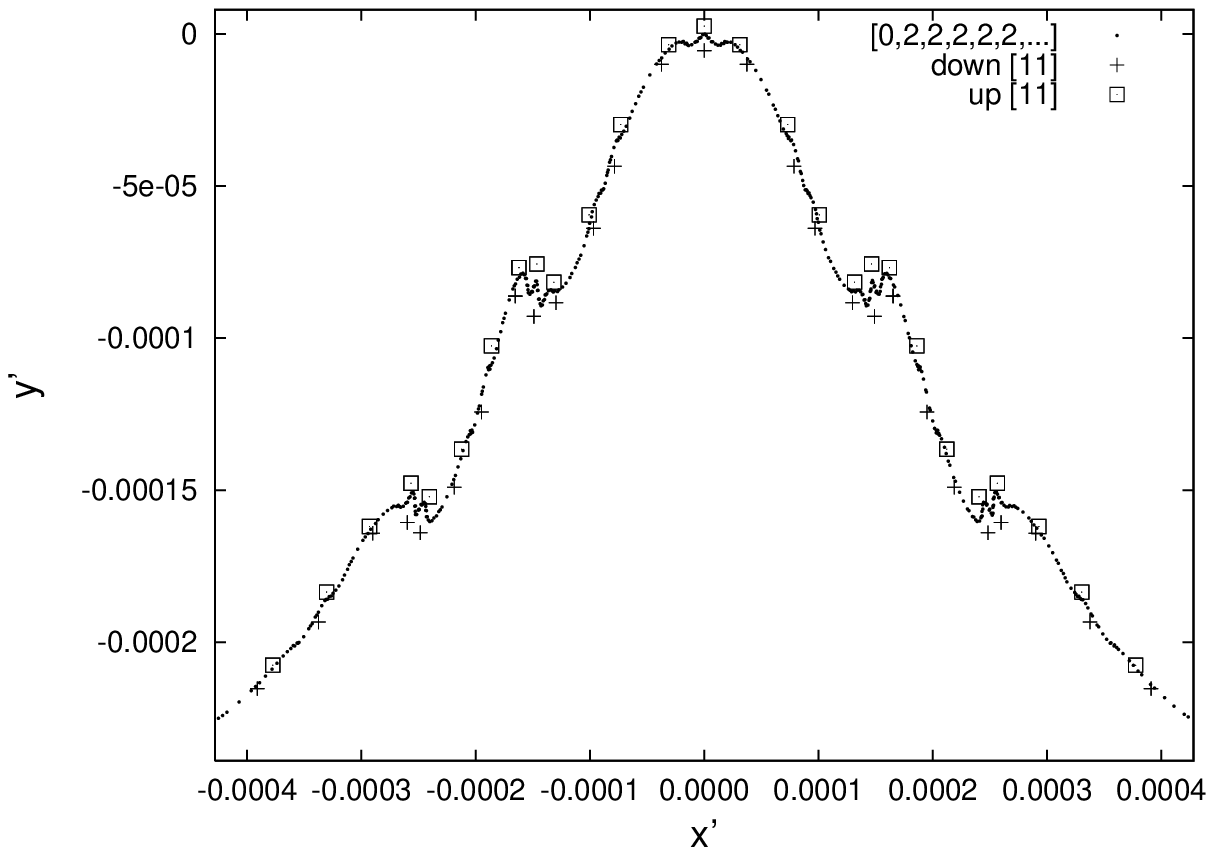}
\caption{Two levels of magnification in symmetry line coordinates
$(x',y')$ of the $\omega=\sqrt{2}-1$ torus at breakup. Also shown
are the nearby up and down orbits of the [7]th (top) and [11]th
(bottom) continued fraction convergents.
  \label{fig:crittorus}}
\end{center} \end{figure}

As in previous studies, the critical torus exhibits invariance under
local re-scaling of the phase space in the neighborhood of the
symmetry lines. Following \rcites{del_castillo97, apte03}, we
compute the scaling factors $\alpha$ and $\beta$ such that the torus
in the vicinity of its intersection with $s_3$ is invariant under
$(x',y') \rightarrow (\alpha^{N} x', \beta^{N} y')$. (The exponent
$N$ is the length of the critical cycle, $N=4$ for
$\omega=\sqrt{2}-1$ and $N=12$ for noble winding numbers.) These
factors are found from the limiting behavior of convergent periodic
orbits: Denoting by $(\hat{x}'_{n,\pm}, \hat{y}'_{n,\pm})$ the
symmetry line coordinates\cite{del_castillo97} of the point on the
up ($+$) or down ($-$) orbit of the $[n]$\/th convergent that is
located closest to $(0,0)$, we compute (see \tabref{crit})
\begin{equation}
\alpha^{N}_{\pm} = \lim_{n\to\infty}
   \left| \frac{\hat{x}'_{n,\pm}}{\hat{x}'_{n+N,\pm}} \right|
   \:, \:\:\:
\beta^{N}_{\pm} = \lim_{n\to\infty}
   \left| \frac{\hat{y}'_{n,\pm}}{\hat{y}'_{n+N,\pm}} \right|
   \:.
\label{eq:alphabeta}
\end{equation}
The scaling invariance of the torus at breakup and of the nearby
periodic orbits is illustrated in \figref{crittorus}.

 As shown in \rcite{del_castillo97}, the numerical data obtained
can be used to compute the unstable eigenvalues, $\delta_1$ and
$\delta_2$, of the RGO ${\mathcal R}$ by

\begin{equation} \frac{1}{\delta_1^N} = \lim_{n\to\infty} \left(
 \frac{b_{[n+N]}\left(a_c\right)-b_c}{b_{[n]}\left(a_c\right)-b_c}\right)
\label{eq:nu1def} \end{equation} and \begin{equation}
\frac{1}{\delta_2^N} = \lim_{n\to\infty} \left(
\frac{a_{c\,[n+N]}-a_c}{a_{c\,[n]}-a_c}\right)\, , \label{eq:nu2def}
\end{equation}
where $a_{c[n]}$ is the $a$ value at which the $\omega$-torus breaks
up along the $b_{[n](a)}$ bifurcation curve.

\begin{table}
\begin{center}\begin{tabular}{|l||l|l|} \hline
\mbox{tail} & $[\ldots,1,1,1,\ldots]$ &$[\ldots,2,2,2,\ldots]$\\
$\mbox{cycle}$ & \quad\qquad 12 & \quad\qquad 4\\
\hline\hline
$\alpha$ & 1.6179 & 2.4725 \\
$\beta$ & 1.6579 & 2.8146 \\
$\delta_1$ & 2.680 & 6.311 \\
$\delta_2$ & 1.584 & 2.455 \\
 \hline \end{tabular} \end{center}
\caption{Universal breakup values for noble tori ($N=12$) and
$\omega=\sqrt{2}-1$ ($N=4$). Values for noble tori are from
\rcite{fuchss06}.} \label{tab:crit}\end{table}

 Our results are displayed in \tabref{crit}. For comparison we
also show the corresponding values for the breakup of tori with
noble numbers.

In summary, we found a new critical cycle of the renormalization
group operator ${\cal R}$, a fixed point of ${\cal R}^4$, which
governs the breakup of the $\omega=\sqrt{2}-1$ shearless invariant
torus. In analogy with the breakup of tori with noble winding
numbers, we expect this result to be the same for all shearless tori
(in the same universality class) with continued fraction expansion
tails consisting of 2's.

\section*{Acknowledgments}

This research was supported by US DOE Contract DE-FG01-96ER-54346.
AW thanks the Dept.\ of Physical and Biological Sciences at WNEC for
travel support.

\end{document}